# Treatment the Effects of Studio Wall Resonance and Coincidence Phenomena for Recording Noisy Speech Via FPGA Digital Filter

Mahmoud I. A. Abdalla

**Abstract**— This work introduces an economic solution for the problems of sound insulation of recording studios. Sound insulation at wall resonance frequency is weak. Instead of acoustical treatment, a digital filter is used to eliminate the effects of wall resonance and coincidence phenomena on recording of speech. Sound insulation of studio is measured to calculate the wall resonance frequency and the coincidence frequency. Pole /zero placement technique is used to calculate the IIR filter coefficients. The digital filter is designed, simulated and implemented. The proposed system is used to treat these problems and it is shown to be effective in recording the noisy speech. In this work digital signal processing is used instead of the acoustic treatment to eliminate the effect of noise at the studio wall resonance. This technique is cheap and effective in canceling the noise at the desired frequencies. Field Programmable Gate Array (FPGA) is used for hardware implementation of the proposed filter structure which provides fast and cheap solution for processing real time audio signals. The implementation is carried out using Spartan chip from Xinlinx achieving higher performance than commercially available software solutions.

◆

## 1 INTRODUCTION

Active noise control (ANC) techniques have attracted much research because they provide numerous advantages over conventional passive methods[1-3]. The maximum external noise spectrum must be reduced to the background noise goal within the space (the studio) by the transmission loss of the walls, window,…..etc. For insulating against outside airborne sounds, general rule is the heavier the wall, the better insulation. The more massive the wall, the more difficult it is for sound waves in air to move in it to and from.

The sound insulation of the studio is weak at the resonance frequency of the wall. To solve this problem at speech recording studio, acoustical treatment must be used. In this work digital signal processing is used instead of the acoustic treatement to eliminate the effect of noise at the studio wall resonance.

Another problem at speech recording studio can be affected by the external noise at the coincidence frequency [4]. When coincidence occurs, it gives rise to a more efficient transfer of energy from the air to the wall to the air on the other side of the wall of the studio. Thus the effective insulation of the wall is lowered producing "coincidence dip" in the insulation curve [4]. Coincidence dip occurs in the frequency range from 1.0 kHz to 4.0 kHz which includes important speech frequencies. In this work digital filter is designed to eliminate the effects of coindence phenomena on the speech rcording environments. This technique is cheap and effective in canceling the noise at the desired frequencies. FPGA is used as the target hardware for the implementation of the filter which has many advantages over other hardware/software platforms. FPGA provides fast parallel processing and the ability to be reconfigured according to variation in system parameters.

## 2 DIGITAL FILTER DESIGN

A filter is essentially a system or network that selectively changes the wave-shap. Digital filter can be used to separate signals that have been combined, such as a musical recording and noise added during the rcording process or to separate signals into their constituent frequencies. Analog filter can be used to accomplish these tasks but digital filters offer greater flexibility and accuracy than analoge filters. Analog filters can be cheaper, faster and have greater dynamic range; digital filterare more flexible than analog filter [5]. The ability to create filters that have arbitrary shape frequency response curve, and filters that meet performance constraint such as bandpass width and transition region width, is well beyond that of analog filters.

Digital filter are broadly divided into two classes, namely infinite impulse response (IIR) and finite impulse response (FIR) filters. The IIR filter equation can be expressed in a recursive form as:

• *Mahmoud I. A. Abdalla is with the Department of electronic and communication, Zagaig University , Zagaig, Egypt.*

$$Y(n) = \sum_{k=0}^{\infty} a_k x(n-k) - \sum_{k=1}^{m} b_k y(n-k) \quad (1)$$

Where the $a_k$ and $b_k$ are the coefficients of the filter. Equatintion (1) can be written in the form of transfer function as:

$$H(z) = \frac{a_0 + a_1 z^{-1} + a_2 z^{-2} + .... + a_n z^{-n}}{1 + b_1 z^{-1} + b_2 z^{-2} + ..... + b_m z^{-m}} \quad (2)$$

It is clear that the current output sample Y(n) is a function of past output as well as present and past input. The choise between FIR and IIR filters depends largely on the relative advantages for the filter types [5]. FIR filter requires more coefficients for sharp cutoff filters than IIR types. Design of a digital filter involes five steps [6] which can be summarized as follows:

1-Specification of the filter requirement:

Requirement specifications include specifying signal characteristics, the characteristics of the filter (desired amplitude and/or phase response), the manner of implementation (as high level language routine or DSP processor-based system), and the cost.

2- Coefficients calculation:

There are many methods to calculate the values of the coefficient $a_k$ or $b_k$ for IIR filter, such that the filter characteristics are satisfied. Impulse invariant method and bilinear transformation as well as poles- zeros placement can be used to calculate the IIR coefficients [7].

3- Representation of a filter by suitable structure:

Realization involves converting a given transfer function into a suitable structure block. Flow diagrams are often used to depict filter structure and they show the computational procedure for implementation of the filter.

4- Analysis of word length effects:

The effects of using a finite number of bits are to degrade the performance of the filter and in some cases to make it unusable. The main sources of performance degradation in digital filters are input/output signal quantization, coefficient quantization, arithmetic roundoff errors and overflow[6,8].

5- Implementation:

The calculated filter coefficients, chosen a suitable structure, verifying that the filter is stable, and the difference equation can be implemented as software routine or in hardware. Whatever the method of implementation, the output of the filter must be computed. To implement a filter basic building blocks are needed:

(i) Memory for storing filter coefficients (ROM).
(ii) Memory (RAM) for storing the present and past inputs and outputs.
(iii) Hardware or software multiplier.
(iv) Adder arithmetic logic units.
(v) Registers(to represent the delay elements).

### 2.1 Pole-Zero Placement

Frequency response of discrete time system can be obtained from Z- transform using several methods [8]. Geometric evaluation of frequency response method is based on its pole-zero diagram. If the transfer function is given by[7]:

$$H(z) = \frac{\prod_{i=1}^{n} K(z - z_i)}{\prod_{i=1}^{m} (z - p_i)} \quad (3)$$

The frequency response is obtaind by substitution with $z = e^{i\omega T}$ in equation (3), where T is the sampling time. A geometric interpretation of Eq.(3) with only two zero and two poles is shown in Fig.1. In this case the frequency response is given by:

$$H(e^{j\omega T}) = \frac{k(e^{j\omega T} - z_1)(e^{j\omega T} - z_2)}{(e^{j\omega T} - p_1)(e^{j\omega T} - p_2)} \quad (4)$$

Or :

$$H(e^{j\omega T}) = \frac{kU_1\angle\Theta_1 U_2\angle\Theta_2}{V_1\angle\Phi_1 V_2\angle\Phi_2} \quad (4)$$

Where $U_1$ and $U_2$ represent the amplitudes of the vectors from the zero to the point $Z=e^{j\omega T}$ and $V_1$ and $V_2$ represent the amplitude of the vector from the poles to the same point as shown in Fig.1. Thus the magnitude and the phase response for the system are:

$$\left| H(e^{j\omega T}) \right| = \frac{U_1 U_2}{V_1 V_2} \quad \text{with k =1} \quad (5)$$

$$\angle H(e^{j\omega T}) = \Theta_1 + \Theta_2 - (\Phi_1 + \Phi_2) \quad (6)$$

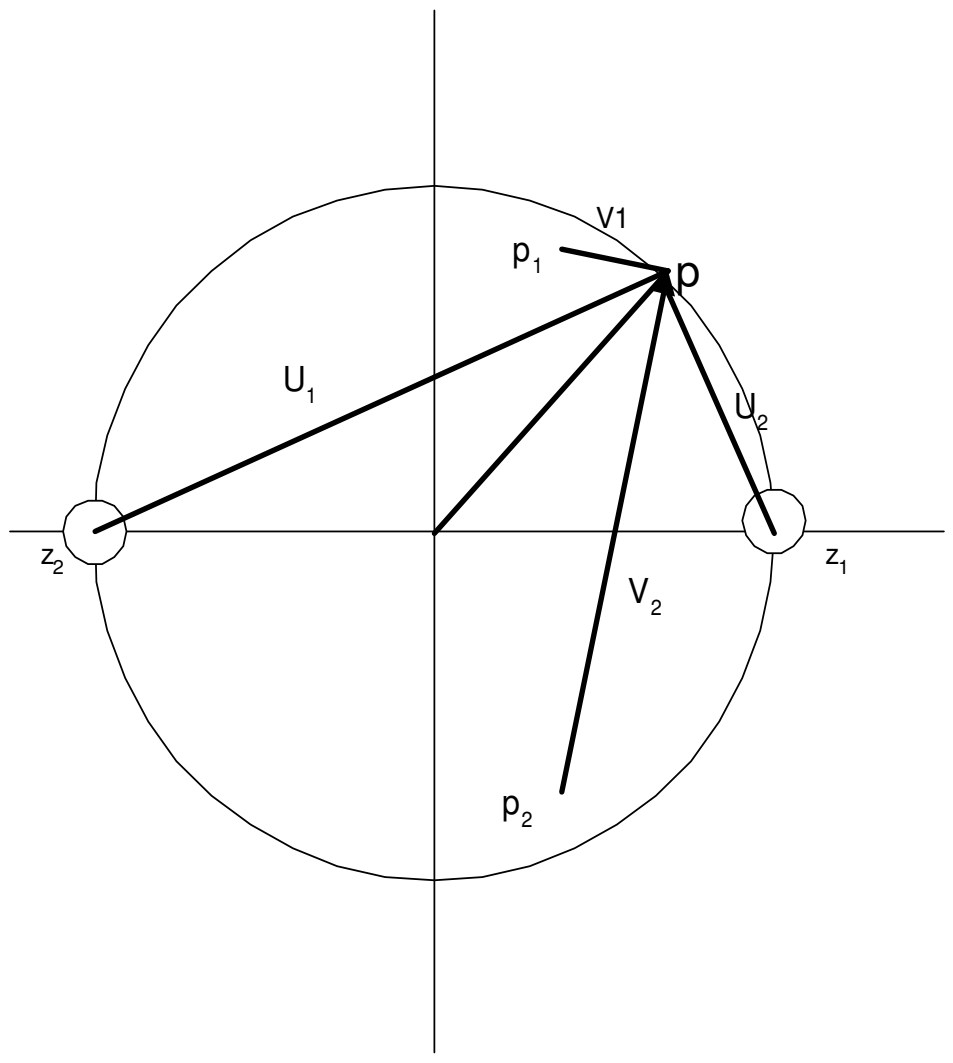

Fig. 1. Geometric evaluation of the frequency response from pole-zero diagram.

The complete frequency response is obtained by evaluating $H(e^{j\omega T})$ as the point P moves from zero to Z= -1. It is evident that as the point P moves closer to $P_1$, the length of the vector $V_1$ decreases and so the magnitude response, $|H(e^{j\omega T})|$, increases. On the other hand, as the point P moves closer to the zero $Z_1$, the zero vector $U_1$ deceases and so the magnitude response, $|H(e^{j\omega T})|$, decreases. Thus at the pole the magnitude response exhibits a peak whereas, at the zero, the magnitude response falls to zero. The design of the proposed filter is based on this principle.

## 3 Experiments

The measurements were carried out at the Building Research Center at Dokki (Giza, Egypt) using Bruel &Kajer instrumentation. The sound source type 4205 generates a steady noise which is filtered in octave band. A rotating microphone boom type 3923 is used to sweep the microphone around a circular path. The sound pressure detected by the microphone is integrated over an appropriated length of time by the building acoustic analyzer 4417. The sound insulation of the walls of the studio is measured. From the sound transmission loss curve, the resonance frequency of studio wall is deduced. The coincidence phenomenon is observed from the same curve. After determining the filter specification, the filter is designed. Measurements were carried to test the system. The noise source is located outside the studio, speech signal is recorded the presence of noise at wall resonance frequency. The filter is implemented using the FPGA and the signal is fed to the computer with load the digital filter to reject the noise. The output of the filter is recorded. The recorded signal is replayed again and compared to the original signal.

## 4 Results And Discussion

The problem is to reject the component of noise at the resonance of the studio walls. So, the resonance frequency of the walls must be calculated. Also the frequency at which the coincidence occurred must be calculated. These frequencies can be observed by investignating the sound insulation of the walls of the studio. The sound reduction (transmission loss) can be calculated from the relation [4], [9]:

$$R = L_1 - L_2 + 10\log_{10}(\frac{S}{A}) \quad \text{dB} \tag{7}$$

Where

$L_1$: average sound pressure level outside the studio.

$L_2$: average sound pressure level in the studio.

S : area of the separate partition.

A : equivalent absorption area in the studio.

The sound absorption area in the studio can be determined from the relation:

$$A = \frac{0.161V}{T} \tag{8}$$

V : volume of the studio

T : reverberation time in the studio

Fig.2 illustrates the measured values of the apparent airborn sound insulation of the studio. The resonance frequency is observed from the curve. It is found to be 315Hz. To reject the component at 315, a pair of complex zeroes are placed on the unit circle corresponding to 315 Hz,that is at angle of 360*315/7400, $\theta_1$= + 15.324324°, $\theta_2$= -15.324324°. The sampling frequency is taken to be 7.4 kHz. To achieve a sharp notch filter and improved amplitude response on either side of the notch frequency, a pair of complex conjugate poles are placed at $r < 1$. The width of the notch is determined by the location of poles at angles of +15.324324° and -15.324324°.

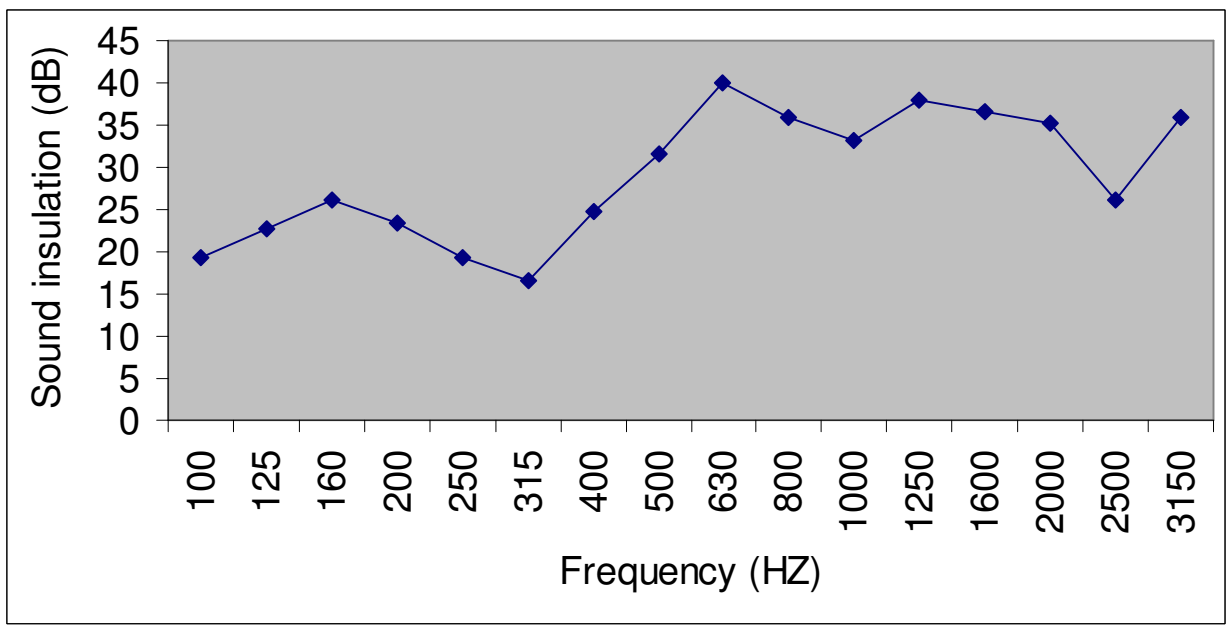

Fig. 2 . Apparent air-born sound insulation of the studio

The poles-zero placement technique is used to calculate the poles and the zero. Substituting with the the zeros and poles in equation (3), the transfer function can be calculated and given by:

$$H_1(z) = \frac{a_0 + a_1 z^{-1} + a_2 z^{-2}}{1 + b_1 z^{-1} + b_2 z^{-2}} \tag{9}$$

With:

$a_0 = 1.000000000000000$ , $b_1 = -1.90960170784598$

$a_1 = -1.92889061398584$ , $b_2 = 0.980100000000000$

$a_2 = 1.000000000000000$

This is the transfer function of the filter to reject the noise at frequency of 315 Hz which is the resonance of the studio walls.

From the sound transmission loss given in Fig.2, it is observed that the coincidence occurred at frequency of 2500 Hz. To reject the nois at this frequency a pair of complex zeroes are placed at an angle of 360*2500/7400, $\theta_1$=121.62°, $\theta_2$=- 121.62°. To achieve a sharp notch filter and improved amplitude response on either side of notch filter, a pair of complex conjugate poles are placed at radius of r<1. The radius r of the pole is determined by the desired bandwidth.An approximated relationship be-tween the radius, r, and the bandwidth,(BW), is given by[6,10]:

$$r = 1 - \frac{BW}{f}\Pi \qquad (10)$$

Where $f$ is the frequency. 'r' in this work is taken to be 0.99 of the unit circle. 'r' is taken 0.99 to locate the poles near the unit circle so that the amplitude of the frequency response is minimum. To eliminate the effect of coincidence phenomena at 2500 Hz, the transfer function of the digital filter is calculated and it has the same form as in equation (9) with :

$a_0 = 1.000000000000000$ , $b_1 = 1.03812842144331$

$a_1 = 1.048614567114460$ , $b_2 = 0.980100000000000$

$a_2 = 1.000000000000000$

This filter rejects the noise at the coincidence frequency.

Finally the two filters are implemented together and the total transfer function is given by:

$$H(z) = H_1(z) * H_2(z) \qquad (11)$$

When the filter order is higher than three, direct realization of the filter is very sensitive to the finite word length effects [11], [14]. The effect of using a finite number of bits is to degrade the performance of the filter and in some cases, it makes the filter unstable. The designer must analyze thes effects and choose suitable word lengths for the filter coefficients. The main sources of performance degradation in digital filter are input/output quantization, coefficient quantization, arithmetic round off errors and overflow which occurs when the result of an operation exceeds permissible word length. The primary effect of quantizing the filter coefficients into a finite number of bits is to alter the positions of the poles and zeros of H(z) in the z-plan. The poles are close to the unit circle so that any significant deviation in them could make the filter unstable.The fewer number of bits used to represent the coefficient, the more deviation will be in pole and zero positions. Large scal overflow occurs at the output of the adders and may be prevented by scaling the the input to the adders in such a way that the outputs are kept low.

The coefficients are quantized using fixed point representation with 1 bit for the integer part and 15 bits for the fraction one yielding 16 bits word length.

The quantized coefficients are:

1- 315 Hz filter:

$a_0 = 1.000000000000000$ , $b_1 = -1.90960693359375$

$a_1 = -1.92889404296875$ , $b_2 = 0.980072021484375$

$a_2 = 1.000000000000000$

2- 2500 Hz filter:

$a_0 = 1.00000000000000$ , $b_1 = 1.03811645507812$

$a_1 = 1.04861450195312$ , $b_2 = 0.980072021484375$

$a_2 = 1.00000000000000$

After quantization the filters are tested for stability and over/under flow using labview 2009 digital filter module[15]. The quantized coefficients errors are minimized with no overflow or underflow arithmeics.

Fig.3 shows the bock diagram of the proposed filter. The frequency response of the filter normalized to the sampling frequency is given in Fig.4. It is clear that the filter rejects the noise speech at 315 and 2500Hz. After implementing the filter, the filter is analzed to ensure its performance.

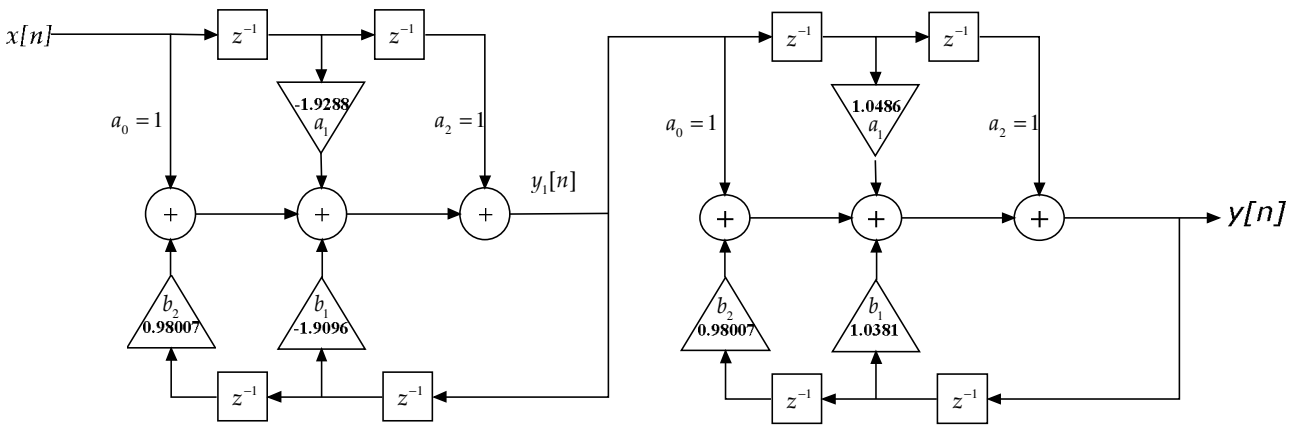

Fig.3. Block diagram representation of the proposed filter

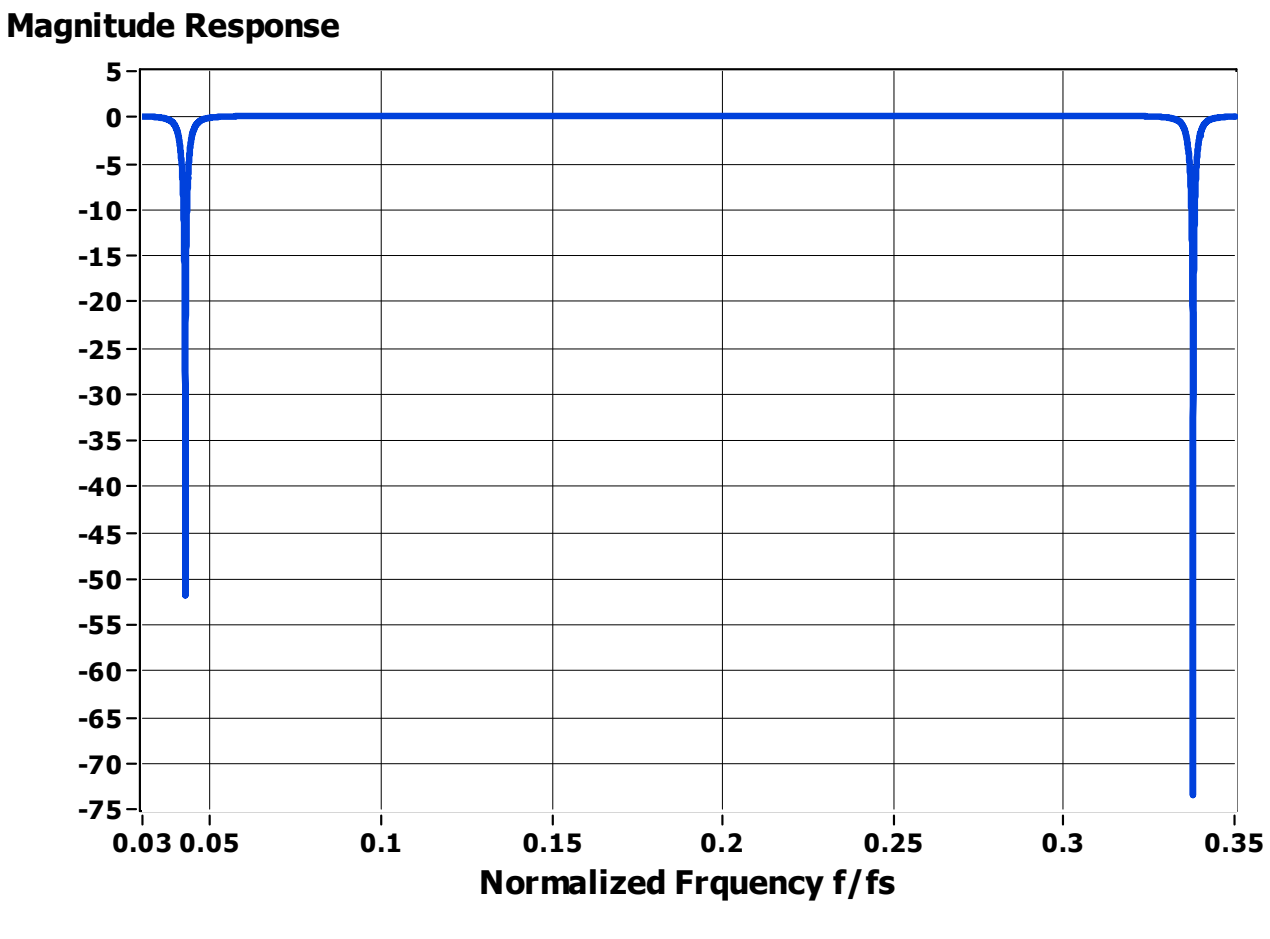

Fig.4. The frequency response of the filter

Fig.5 shows poles and zeros of the filters. It is clear from zooming of Fig.5 that the poles are inside the unit circle while the zeros are on the unit circle. Fig.6 illu-strates that the zeros are on the unit circle while the poles are inside the unite circle.

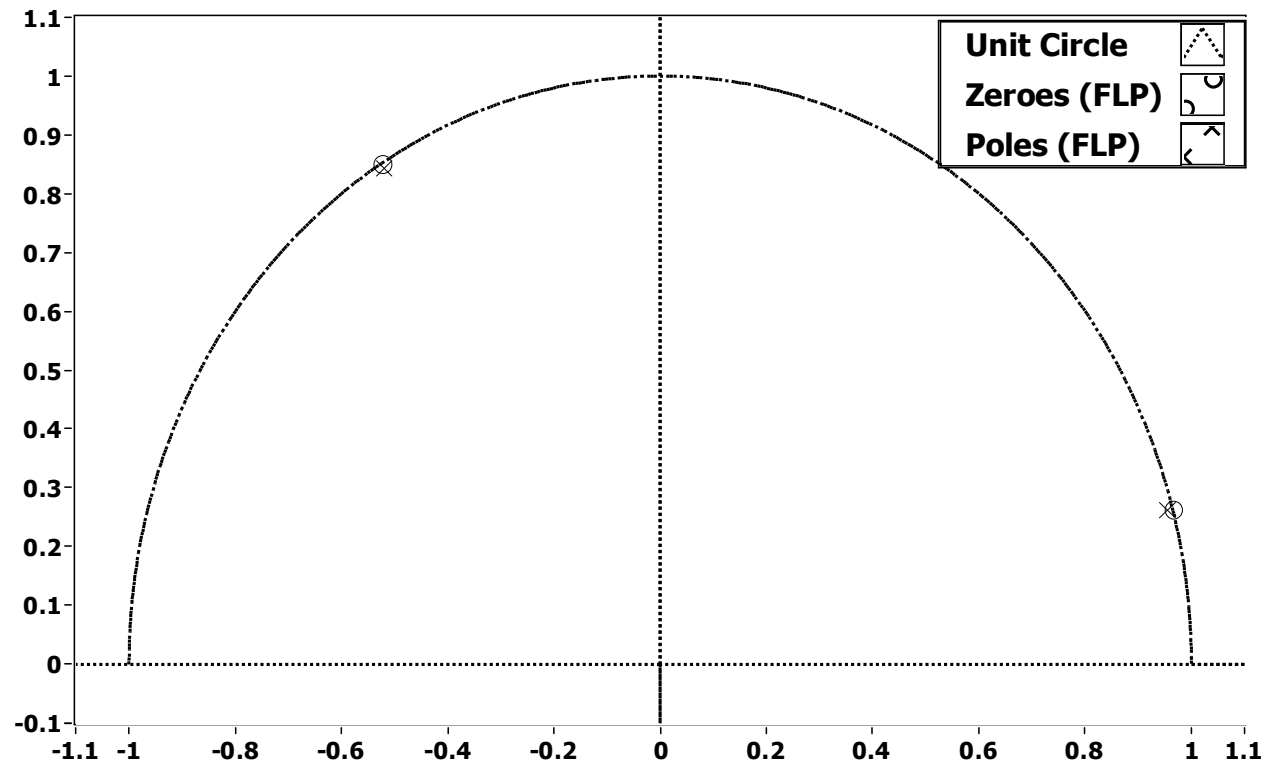

Fig.5. Location of poles and zeros of the filters

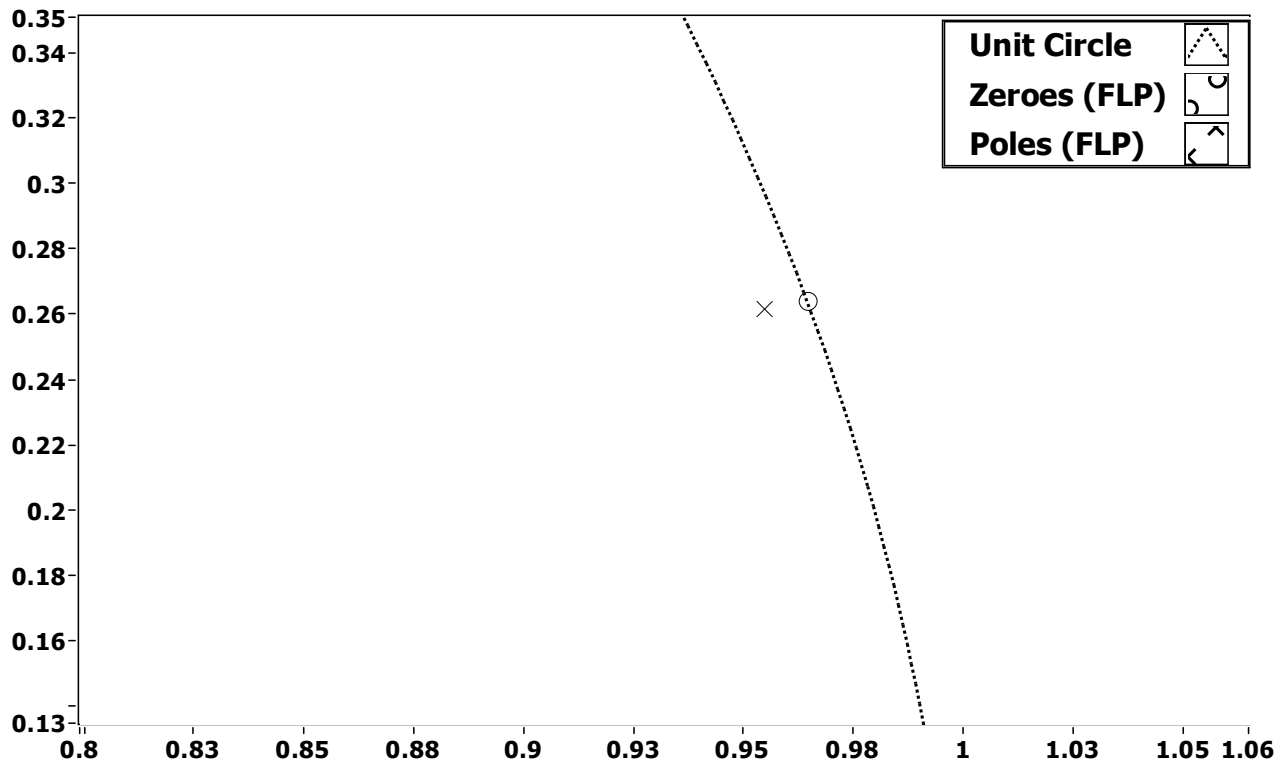

Fig.6-a. Location of the first pole and zero after normalization

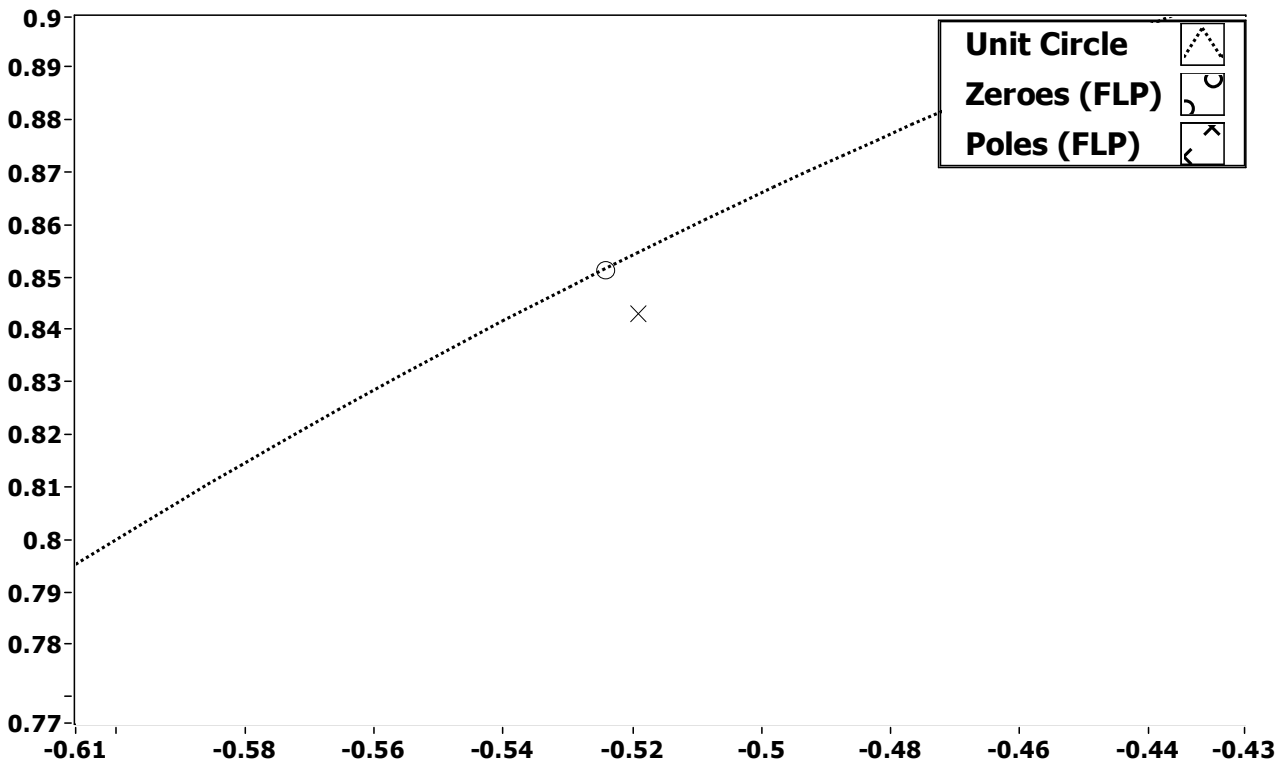

Fig.6-b. Location of the second pole and zero after normalization

## 5 FPGA IMPLEMENTATIONS:

The time domain equation of the proposed IIR filter is given by:

$$y[n] = a_0 x[n] + a_1 x[n-1] + a_2 x[n-2] + b_1 y[n-1] + b_2 y[n-2] \quad (12)$$

A simplified expression for the filter equation is chosen to help in simplifying FPGA implementation by setting $a_0$ and $a_2$ to 1. This simplification results in increasing the gain to a value slightly greater than 1 which is acceptable in audio processing. The simplified equation is given by:

$$y[n] = x[n] + a_1 x[n-1] + x[n-2] + b_1 y[n-1] + b_2 y[n-2] \quad (13)$$

This equation needs 3 multipliers and a 5 input adder circuit which is more efficient than what we need if we use the design with equation (12).

The coefficients are all quantized in 16 bits fixed point representation. The samples of the signal are represented by 8-bits. For the implementation using FPGA, we need 3 multipliers 8*16 bits, which will consume very large cells and the implementation cost will be increased.

So the multiplication can be done using a lookup tables implemented in a memory outside the FPGA, and hence the required operations needed from the FPGA are the shift and accumulate to compute the next sample value as shown in Fig 7.

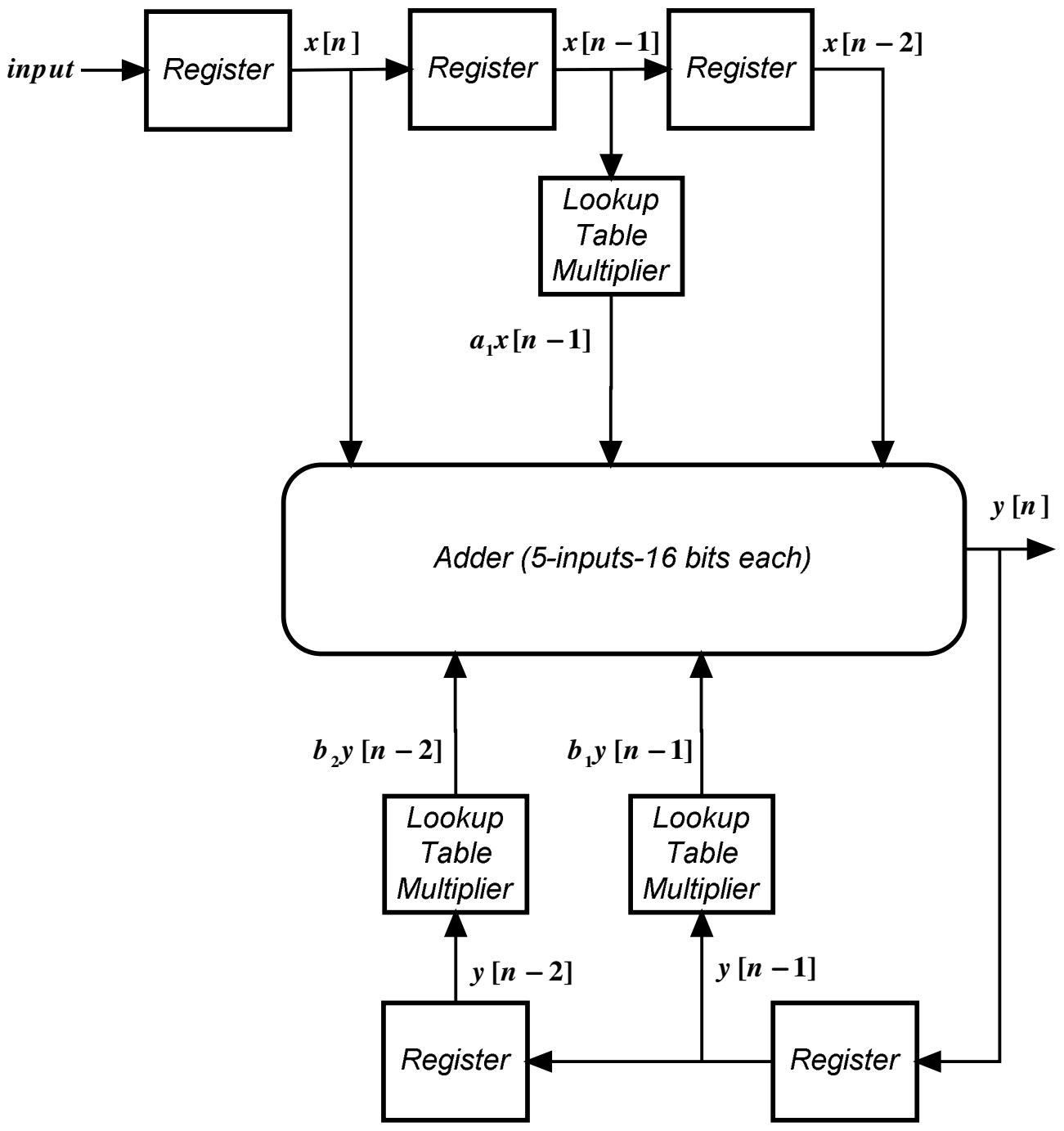

Fig. 7 FPGA implementation of one section of the IIR filter

We use clocked registers to store the values of $x[n]$, $x[n-1]$, $x[n-2]$, $y[n-1]$, $y[n-2]$ and a 16-bits adder circuits that can accumulate 5 inputs to produce $y[n]$.

Using this technique minimizes cells usage of the FPGA and makes the implementation visible using low cost FPGA chips.

The same block as in Fig. 7 is used to implement the 2nd section of the filter at 2500 Hz.

Table 1 summarizes FPGA utilization using Spartan chip from Xilinx[16] in terms of the CLB (Configurable Logic Block) and maximum operating frequency.

TABLE 1 FPGA PERFORMANCE USING SPARTAN CHIP

| Implemented Filter Type | Performance | | |
|---|---|---|---|
| | CLBs | Maximum Frequency | Number of Flip -Flops |
| 315 Notch filter | 182 | 23.1 MHz | 198 |
| 2500 Notch filter | 182 | 23.1 MHz | 198 |
| Cascaded Filters | 365 | 20.5 MHz | 377 |

The same filters are designed using labview with FPGA module as the target hardware and the results are shown in Table 2.

TABLE 2. FPGA PERFORMANCE USING COMMERCIAL SOFTWARES

| Implemented Filter Type | Performance | | |
|---|---|---|---|
| | CLBs | Maximum Frequency | Number of Flip -Flops |
| 315 Notch filter | 722 | 2.9 MHz | 732 |
| 2500 Notch filter | 722 | 2.9 MHz | 732 |
| Cascaded Filters | 1531 | 1.87 MHz | 1509 |

The obtained result introduces a performance advantage over professional software packages such as labview or matlab which are restricted to expensive DSP specific FPGA board. Also the maximum sampling frequency is much greater than what obtained by labview tools.

After designing the filter, the system is tested experimentally. The noise source is located outside the studio.The speech signal is recorded with the presence of the filter and without it. Fig.8 and Fig.9 show the spectrum of each signal.

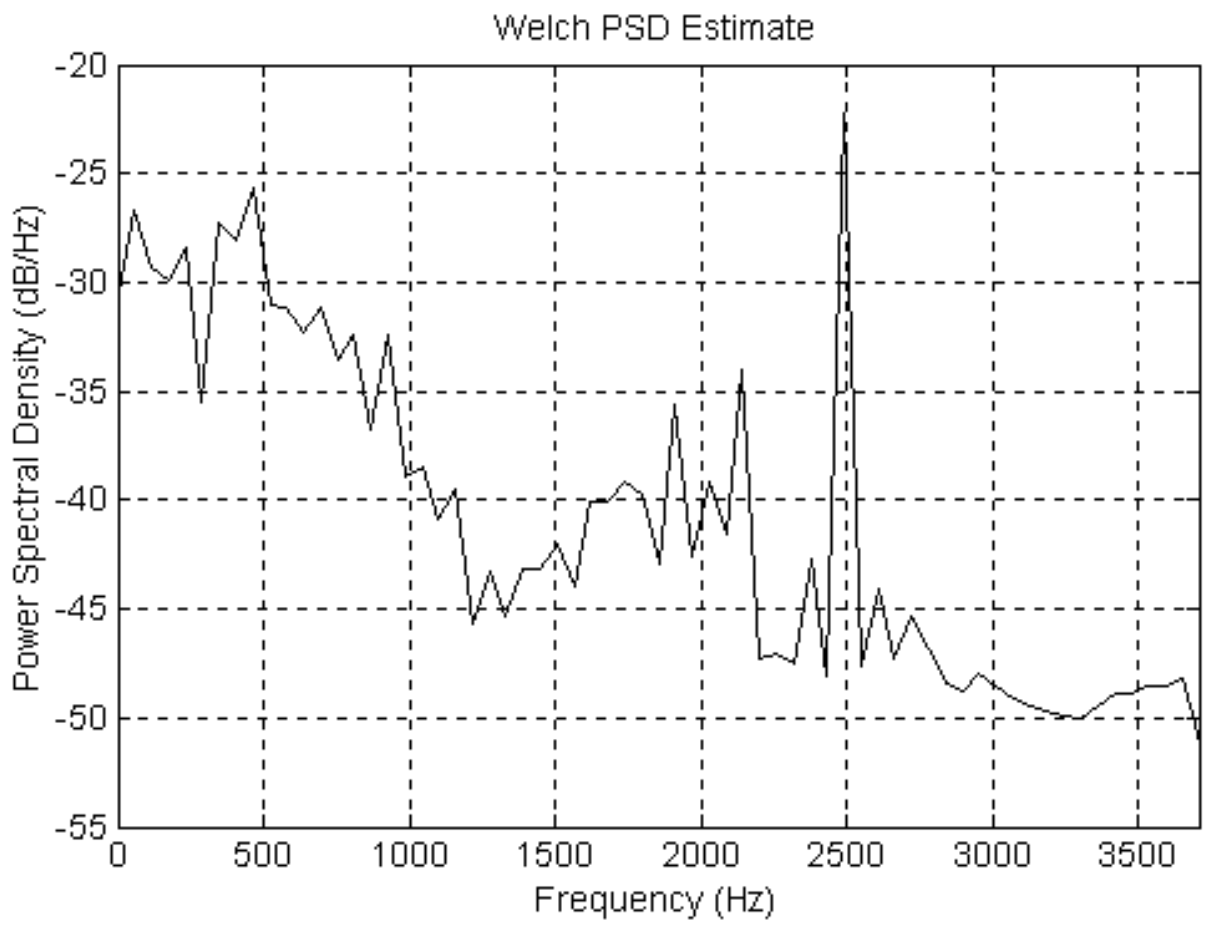

Fig.8. The recorded speech signal without using the filter

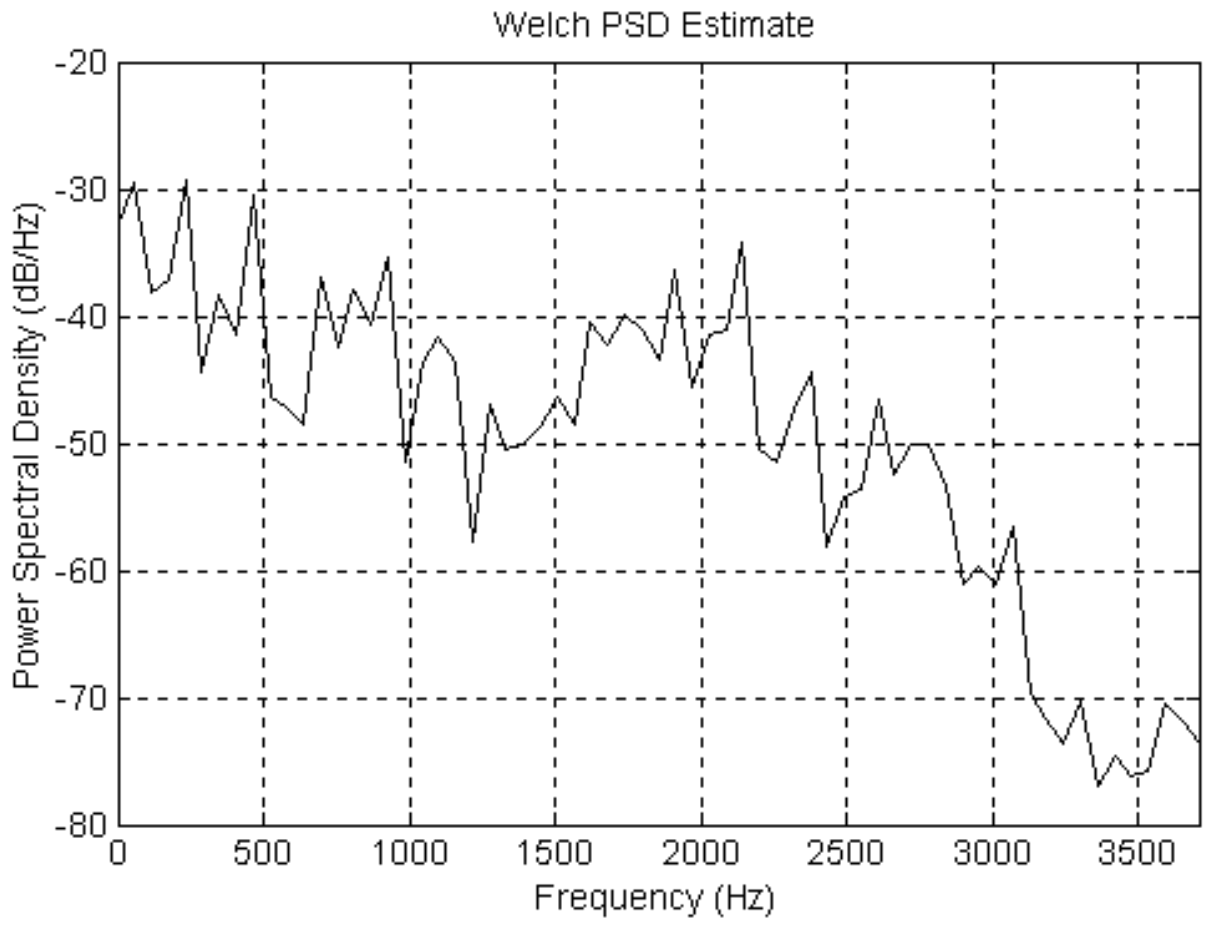

Fig.9. The recorded speech signal with the filter

## 6 CONCLUSION

An economic technique is introduced and applied successfully to eliminate the effect of wall resonance and coincidence phenomena on the sound insulation. The proposed technique is designed, simulated, and tested. The experiments show that this method can be applied in the case of recording noisy speech in studio especially after building the studio instead of acoustical treatment. An efficient FPGA implementation of the proposed filter is designed and implemented using low cost FPGA chips.